\documentclass[conference]{IEEEtran}
\usepackage{graphicx}
\usepackage{epsfig}
\usepackage{amsmath}
\pdfoutput=1

\usepackage{graphicx}
\begin{document}

\title{Network Coding-based Cooperative ARQ Scheme}

\author{\IEEEauthorblockN{Angelos Antonopoulos and Christos Verikoukis}
\IEEEauthorblockA{Telecommunications Technological Centre of Catalonia (CTTC)\\
Castelldefels, Barcelona, Spain\\
Email: \{aantonopoulos, cveri\}@cttc.es}\\
}

\maketitle              % typeset the title of the contribution

\begin{abstract}
%\boldmath
In this paper we introduce a novel Automatic Repeat reQuest (ARQ) scheme for cooperative wireless networks. Our scheme adopts network coding techniques in order to enhance the total bandwidth of the network by minimizing the total number of transmissions. The performance of the proposed approach is evaluated by means of computer simulations and compared to other cooperative schemes, while an analytical solution is provided to validate the results.
\end{abstract}

\begin{IEEEkeywords}
Network Coding; Cooperative Networks; Medium Access Control (MAC); Automatic Repeat reQuest (ARQ).
\end{IEEEkeywords}

\IEEEpeerreviewmaketitle

\section{Introduction}
\label{sec:intro}

\IEEEPARstart{W}{ireless} communication has experienced an impressive growth during the last decades. Lately, new techniques such as cooperation among nodes and network coding have been introduced in order to improve the network's performance and provide the communication with robustness, diversity, higher data rates and security. These new technologies generate the need of designing new Medium Access Control (MAC) protocols that exploit the benefits of the aforementioned techniques in order to efficiently use the network resources.

The concept of cooperation was introduced by Cover et al. \cite{IEEEhowto:coop} in their fundamental paper on relay channels. Their work analyzed the capacity of the three-node network consisting of a transmitter, a receiver and a partner (relay)\footnote[1]{Note that the words ``partner", ``relay" and ``helper" are used interchangeably in this paper.}. In their model, the spatial diversity gain is obtained by exploiting different channels seen by different nodes for transmitting data.

On the other hand, network coding is an area that has emerged in 2000 \cite{IEEEhowto:nc1}-\cite{IEEEhowto:nc2}, and since then has attracted an increasing interest, as it promises to have a significant impact in both theory and practice of networks. We can broadly define network coding as allowing intermediate nodes in a network to not only forward but also process the incoming information flows. Most of the work on this topic focuses on the physical layer aspect \cite{IEEEhowto:phync1}-\cite{IEEEhowto:phync4} while only few works examine these techniques considering the MAC layer effect \cite{IEEEhowto:macnc1}-\cite{IEEEhowto:macnc3}.

The main contribution of our proposed scheme lies on the fact that we combine both cooperative and network coding techniques in order to enhance the system's performance. To the best of our knowledge, there is no proposed MAC scheme in the literature that implements network coding in cooperative ARQ schemes, while there is a limited number of papers that apply network coding in cooperative schemes that take advantage of the multi-rate capability of wireless standards \cite{IEEEhowto:mr1}.

The rest of the paper is organized as follows. Section \ref{sec:rw} gives the basic background on cooperative networking and outlines the related work on MAC layer protocols for both simple and network coding-based cooperative schemes in the literature. In Section \ref{sec:ncc} we introduce our proposed protocol NCC-ARQ along with a brief mathematic analysis. The validation of the the analytical model and the numerical results are provided in Section \ref{sec:results}. Finally, Section \ref{sec:conclusion} concludes the paper.

\section{Background and Related Work}
\label{sec:rw}

\subsection{Cooperative Communication}

In the context of cooperative communications, several schemes focused on MAC layer aspect have been already proposed in literature \cite{IEEEhowto:maccoop1}-\cite{IEEEhowto:macmr4}. These works can be classified into two main categories: i) the cooperative ARQ-based protocols and ii) the protocols that transform one-hop transmissions to multi-hop transmissions by exploiting the multi-rate capabilities of the wireless systems.

\subsubsection{Cooperative ARQ-based protocols}
Forward Error Correction (FEC) and Automatic Repeat reQuest (ARQ) algorithms are two basic error control methods for data communications \cite{IEEEhowto:arq}. ARQ schemes have received considerable attention for data transmissions due to their simplicity and higher reliability, compared to FEC schemes.

Regarding the protocols falling in this category \cite{IEEEhowto:maccoop1}-\cite{IEEEhowto:maccoop3}, the retransmissions are initiated by the destination after an erroneous packet reception. The helpers in a network are enabled to relay the original packets to a specific destination, as ARQ defines, using higher data rates or better channel conditions in terms of Signal to Noise Ratio (SNR) values.

\subsubsection{Protocols that transform one-hop transmissions to multi-hop transmissions}

By using the concept of adaptive modulation \cite{IEEEhowto:am}, mobile stations in a multi-rate wireless network assign the modulation scheme and transmission rate according to the detected Signal-to-Noise ratio (SNR) and the required transmission quality. Each modulation scheme could be further mapped to a range of SNR in a given transmission power. To achieve high transmission efficiency in wireless systems, stations select the highest available rate modulation scheme according to the detected SNR.

The protocols of this class \cite{IEEEhowto:macmr1}-\cite{IEEEhowto:macmr4} transform single one-hop transmissions to multi-hop transmissions according to the channel conditions. Specifically, when the channel state between the relay and the destination is better than the channel between the source and the destination, a two-hop transmission is preferred instead of the direct transmission.

\subsection{Cooperation and Network Coding}

Last years, there is a trend towards using network coding in cooperative communications. The initial attempts for developing network coding-based cooperative communications focused on physical layer schemes \cite{IEEEhowto:phycoopnc1}-\cite{IEEEhowto:phycoopnc3}. These approaches refer to the coding gain and optimal power allocation in simple cooperative topologies, usually considering one relay or cooperation among the users.

However, the innovation of using network coding in cooperative communications is not confined only in the physical layer. Tan et al. \cite{IEEEhowto:mr1} presented one of the few works that focus on MAC layer aspect of network coding-based cooperative communication. Their proposed protocol, called CODE, exploits the benefits of both network coding and multi-rate capability of IEEE 802.11 Standard. Specifically, the coding of the packets takes place at the relay nodes, under two basic conditions: i) the direct link between the sender and the receiver is poor and exists one or more relay candidates that experience better link conditions and ii) the traffic is bidirectional.
	
\section{Proposed Network Coding-based Cooperative ARQ Scheme}
\label{sec:ncc}

\subsection{Motivation}

The main motivation for this paper lies on exploiting the advantages of both cooperative communication and network coding. The limited number of work found in the literature related to applying network coding in cooperative schemes has led us to implement a new MAC protocol, called Network Coding-based Cooperative Automatic Repeat reQuest (NCC-ARQ), exclusively designed for cooperative networks. The cooperation during the operation of the protocol results in a distributed cooperative ARQ scheme, while network coding techniques are used in order to improve the performance of the system.

\begin{figure}[htb]
\centering
\includegraphics[scale=0.3]{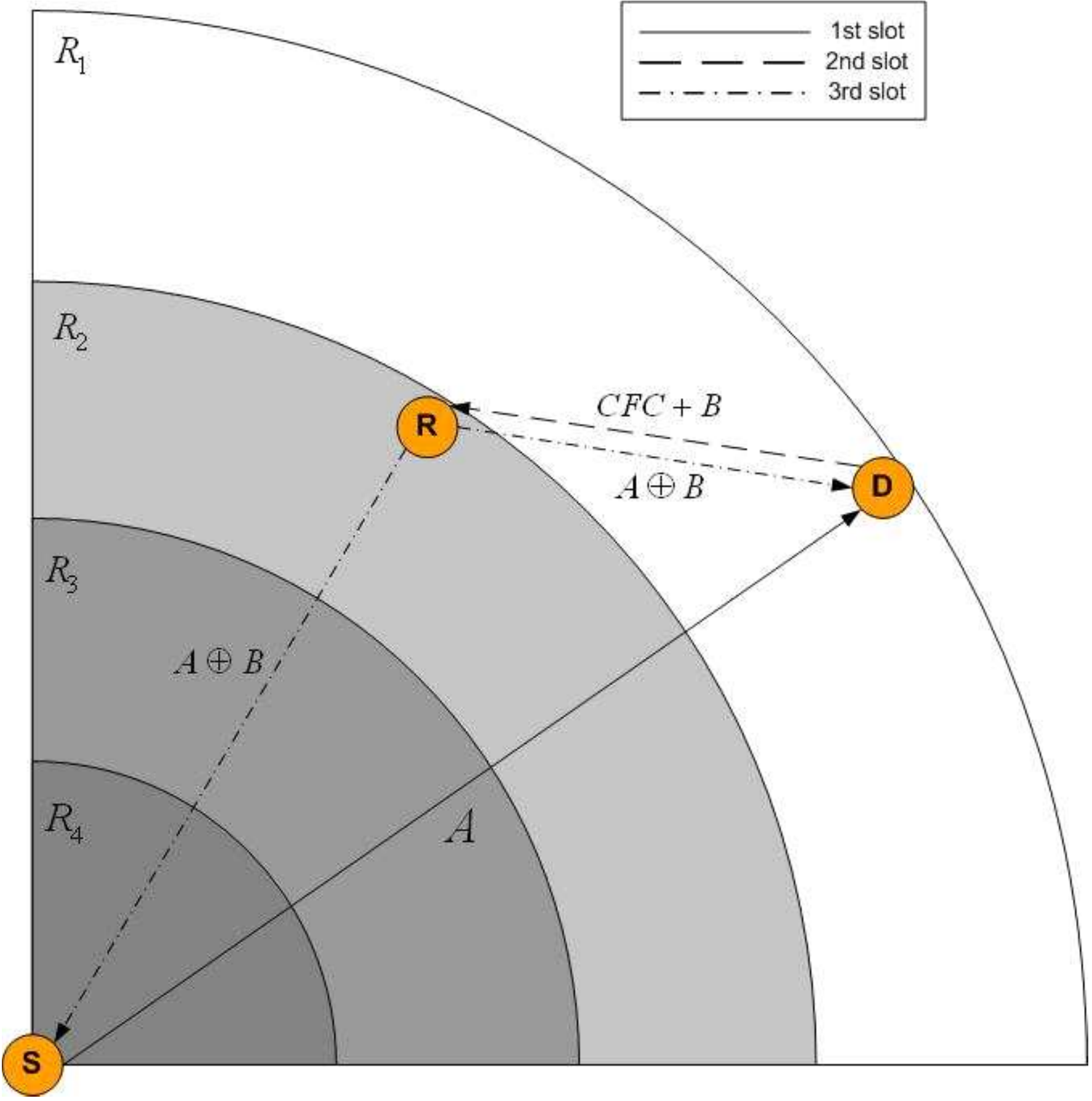}
\caption{General idea for NCC-ARQ scheme}\label{f1}
\end{figure}

The main design goal of NCC-ARQ is twofold: i) to enable the IEEE 802.11 stations to request their neighborhood to cooperate upon the erroneous reception of a data packet and ii) to allow partner nodes to code the data packets to be transmitted before relaying them. The aforementioned goals will be further clarified once we will have described the operation of the protocol. Following, the operation of the protocol is explained in detail, while a simple scenario subjects to the initial principles of NCC-ARQ is depicted in Figure \ref{f1}.

\subsection{Protocol Description}

When NCC-ARQ is applied in the network, all the nodes should operate in promiscuous mode in order to be able to capture all ongoing transmissions and cooperate, if required. In addition, they should keep a copy of any received data packet (regardless of its destination address) until it is acknowledged by the destination station.

Whenever a data packet is received with errors at the destination node, a cooperation phase can be initiated. The error control could be performed by checking a cyclic redundancy code (CRC) attached to the header of the packet or any other equivalent mechanism. The cooperation phase is initiated by the destination station by transmitting a Call for Cooperation (CFC) message to the best relay\footnote[2]{Note that the most appropriate relay selection algorithm for our scheme is the one proposed by Li et al. \cite{IEEEhowto:selection}, which considers the design of joint network coding and relay selection in two-way relay channels.} in terms of channel conditions (i.e. SINR) after sensing the channel idle for a Short Interframe Space (SIFS) period. This message has the form of a control packet and higher priority over regular data traffic, since data transmissions in IEEE 802.11 take place after a longer period of silence (DIFS). Furthermore, in the special but not rare case of bidirectional traffic, when the destination station has a data packet for the source station, it transmits this packet piggybacked with the CFC message.

Upon the reception of the CFC, the helper node gets ready to forward its information. Since the relay has already stored the packets that destined both to the destination (the cooperative packet) and to the source (the piggybacked packet), it creates a new coded packet by combining the two existing data packets, using the XOR method. In this point we have to state that NCC-ARQ uses the same frame structure and follows the same principles with the IEEE 802.11 Standard, thus maintaining the backwards compatibility with it. However, there have been some modifications that are necessary in order for the protocol to exploit efficiently the advantages of using both cooperative and network coding techniques:
\begin{enumerate}
\item There is no expected ACK packet associated to the data packets that are sent piggybacked with the CFC message.
\item In case of bidirectional traffic, the packet that is destined back to the source is sent along with the Call for Cooperation packet, without taking part in the contention phase.
\item There are ACK packets for the multicast transmission of the coded packet in order to provide a reliable communication scheme.
\end{enumerate}

Once the source and the destination receive the network coded packet from the relay, they are able to decode it and extract the respective original data packets. Subsequently, they acknowledge the received data packet by transmitting the respective ACK, thus terminating the cooperation phase. In case that the received coded packets could not be decoded  after a certain maximum cooperation timeout due to transmission errors, the relay is obliged to forward again the network coded packet.

\subsection{Operational Example}

\begin{figure}[htb]
\centering
\includegraphics[scale=0.50]{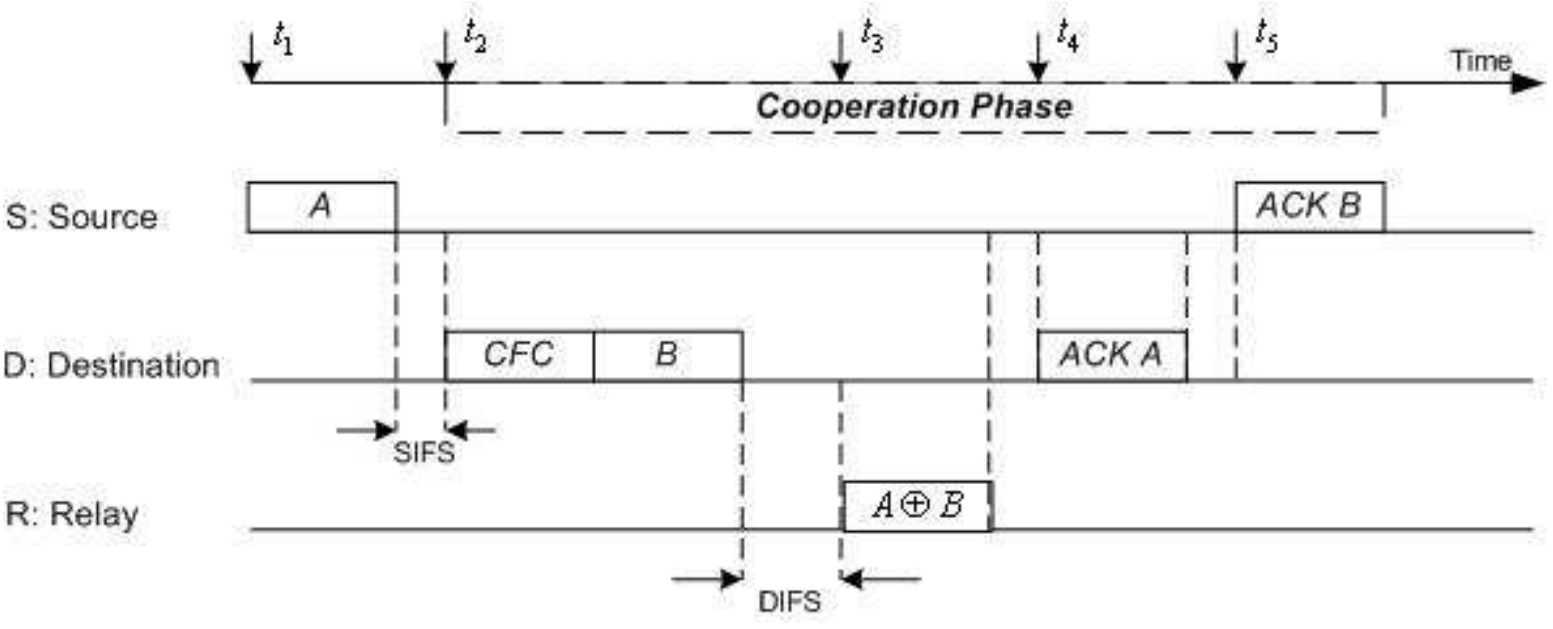}
\caption{NCC-ARQ example of operation}\label{f2}
\end{figure}

In this subsection we provide a simple example in order to clarify the operation of the protocol. A basic network topology with 3 stations is considered, all of them in the transmission range of each other. A source station ($S$) transmits a data packet ($A$) to a destination station ($D$) which has also a packet ($B$) destined to the source station. There is also one relay ($R$) that has been chosen as the most appropriate helper node in terms of Signal to Interference-Noise Ratio (SINR) and supports this particular bidirectional communication. The whole procedure is depicted in Figure \ref{f2} and explained as follows:
\begin{enumerate}
\item At instant $t_{1}$, station $S$ sends the data packet $A$ to station $D$.
\item Upon reception, at instant $t_{2}$, station $D$ fails to demodulate the packet $A$, thus transmitting a $CFC$ packet to $R$ along with the data packet $B$, destined to the station $S$.
\item At instant $t_{3}$, the relay $R$  transmits the coded packet $A \oplus B$ to the nodes $S$ and $D$ simultaneously.
\item At instant $t_{4}$  the station $D$ sends back an $ACK$ packet since it is able to decode properly the XOR-ed packet and retrieve the original packet $A$.
\item At instant $t_{5}$  the node $S$ acknowledges the packet $B$ since it is able to decode properly the coded packet $A\oplus B$.
\end{enumerate}

\subsection{Protocol Analysis}
\subsubsection{Delay Analysis}

The total time that is required in order for two packets to be exchanged in NCC-ARQ can be computed as:
\begin{equation}
\label{eq:delay}
 	E[D]= E[T_{A}]+E[T_{COOP}]
\end{equation}

where $E[T_{A}]$ denotes the average time for a transmission of a single data packet between the source and the destination, while $E[T_{COOP}]$ corresponds to the average time required for the cooperative transmission scheme to be completed.

Furthermore, $E[T_{A}]$ is directly correlated with the network configuration, while $E[T_{COOP}]$ represents the average delay of the cooperation phase due to the contention among the nodes and the number of required retransmissions, as well. In this paper we assume that the retransmissions take place using always the best relay in term of channel conditions, thus eliminating the need of contention among the relays. Hence, the term $E[T_{COOP}]$  can be analytically expressed as
\footnotesize
\begin{equation}
\label{eq:coop}
\begin{aligned}
 &E[T_{COOP}]=T_{SIFS}+T_{CFC}+T_{B}+T_{DIFS}+T_{ONC}+\\
 &+E[r]\cdot T_{A \oplus B}+T_{SIFS}+T_{ACK}+T_{SIFS}+T_{ACK}
 \end {aligned}
\end{equation}

\normalsize

In the above expression, $T_{SIFS}$ and $T_{DIFS}$ represent the duration of $SIFS$ and $DIFS$ waiting times, respectively. $T_{CFC}$, $T_{ACK}$ and $T_{B}$ denote the transmission times of the packets $CFC$, $ACK$ and $B$, respectively. Furthermore, $T_{A \oplus B}$ represents the transmission time for a network coded packet, while $T_{ONC}$ is the overhead time that a relay needs in order to perform the network coding techniques. $E[r]$ is the average number of the retransmissions that are required in order to properly decode the X-OR packets at the destination nodes. It depends on the channel conditions and specifically on the packet error rate ($PER$) between the relay and the destination nodes. Lower values of $PER$ imply higher probability for successful decoding of the packets at the destination nodes. This relationship could be mathematically expressed as:

\begin{equation}
\label{eq:retr}
 	E[r]=1/(1-PER_{R\rightarrow D})
\end{equation}

\subsubsection{Throughput Analysis}

Two packets are delivered simultaneously to the respective destination nodes by applying network coding techniques in each transmission cycle of NCC-ARQ. Thus, the system's aggregated throughput can be defined as:

\begin{equation}
\label{eq:throughput}
 	S[b/s]= 2 \cdot \frac{E[P]}{E[D]}
\end{equation}

where $E[P]$ represents the average packet payload, while $E[D]$ derives by the formulas~(\ref{eq:delay}) and (\ref{eq:coop}) and has already been analyzed in the previous subsection. Furthermore, the coefficient 2 in formula~(\ref{eq:throughput}) is used to express that two packets are delivered in each transmission.

\section{Performance Results}
\label{sec:results}

In order to evaluate the performance of the NCC-ARQ we have developed an event-driven C++ simulator that executes the rules of the protocol. In this section we present the simulation set up and results of our experiments.

\subsection{Simulation Scenario}

We simulate an 802.11g network formed by a pair of transmitter-receiver (the two nodes are both transmitting and receiving data) and a relay node that facilitates the communication, all of them in the transmission range of each other. Furthermore, the relay node is able to perform network coding to its buffered packets before relaying them. In order to focus on the analysis of the impact of both network coding and cooperative communication, the following assumptions have been made:
\begin{enumerate}
\item The traffic is bidirectional, i.e. the destination node has always a packet destined back to the source node.
\item Original transmissions from source to destination are always received with errors, thus initiating a cooperative phase.
\item The channel between the source and the destination is error symmetric, i.e. $PER_{S\rightarrow D} =PER_{D\rightarrow S}$
\item The packet error rate ($PER$) - and consequently the  required number of retransmissions ($E[r]$) that have to be made by the relay until the packets received correctly - is known a priori.
\end{enumerate}

The configuration parameters of the stations in the network are summarized in TABLE I considering the IEEE 802.11g PHY layer \cite{IEEEhowto:80211g}. Furthermore, the time for applying network coding $T_{ONC}$ is considered to be negligible, since the coding takes place between only two packets.

\begin{table}[h]
\caption{System Parameters} \label{t1}
\begin{center}
\begin{tabular}{|c|c|c|c|}
\hline
\textbf{Parameter} & \textbf{Value} & \textbf{Parameter} & \textbf{Value} \\ \hline
\textit{MAC Header} & 34 bytes & \textit{DATA packets} & 1500 bytes \\ \hline
\textit{PHY Header} & 96 $\mu$sec & \textit{SIFS} & 10 $\mu$sec \\ \hline
\textit{ACK, CFC} & 14 bytes & \textit{DIFS} & 50 $\mu$sec \\ \hline
\textit{Source Control Rate} & 6Mb/s & \textit{Source Data Rate} & 6Mb/s \\ \hline
\textit{Relay Control Rate} & 6Mb/s & \textit{Relay Data Rate} & 54Mb/s \\ \hline
\end{tabular}
\end{center}
\end{table}

\begin{figure}[htb]
\centering
\includegraphics[scale=0.65]{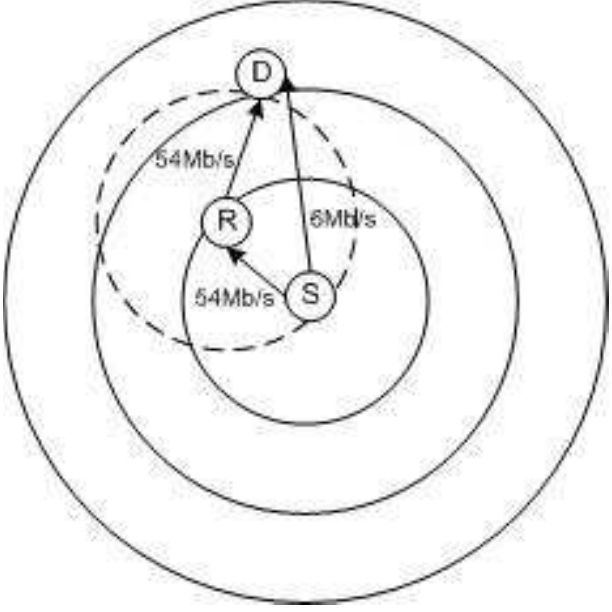}
\caption{Simulation Scenario}\label{f3}
\end{figure}

The simulation scenario, which is depicted in Figure \ref{f3}, shows the topology of the network. Apart from the assumptions that have already been mentioned, we further assume that the transmission data rates between the relay and the source or the destination is 54 Mb/s, as the relay is located close to both nodes, while the data rate between the source and the destination is 6 Mb/s. The network operates under saturated conditions, which means that the nodes have always packets to send in their buffers.

In order to evaluate our approach, we compare our scheme with a simple cooperative ARQ scheme (C-ARQ), where the bidirectional communication takes place in two steps. In the first step, node $S$ sends the packet to $D$ and, upon the erroneous reception, $D$ transmits the $CFC$ packet, thus triggering the relay to retransmit the packet. In the second step, node $D$ transmits its own packet to $S$ and the same procedure as in the first step is repeated, thus consuming valuable network resources.

\subsection{Simulation Results}

\begin{figure}[htb]
\centering
\includegraphics[scale=0.55]{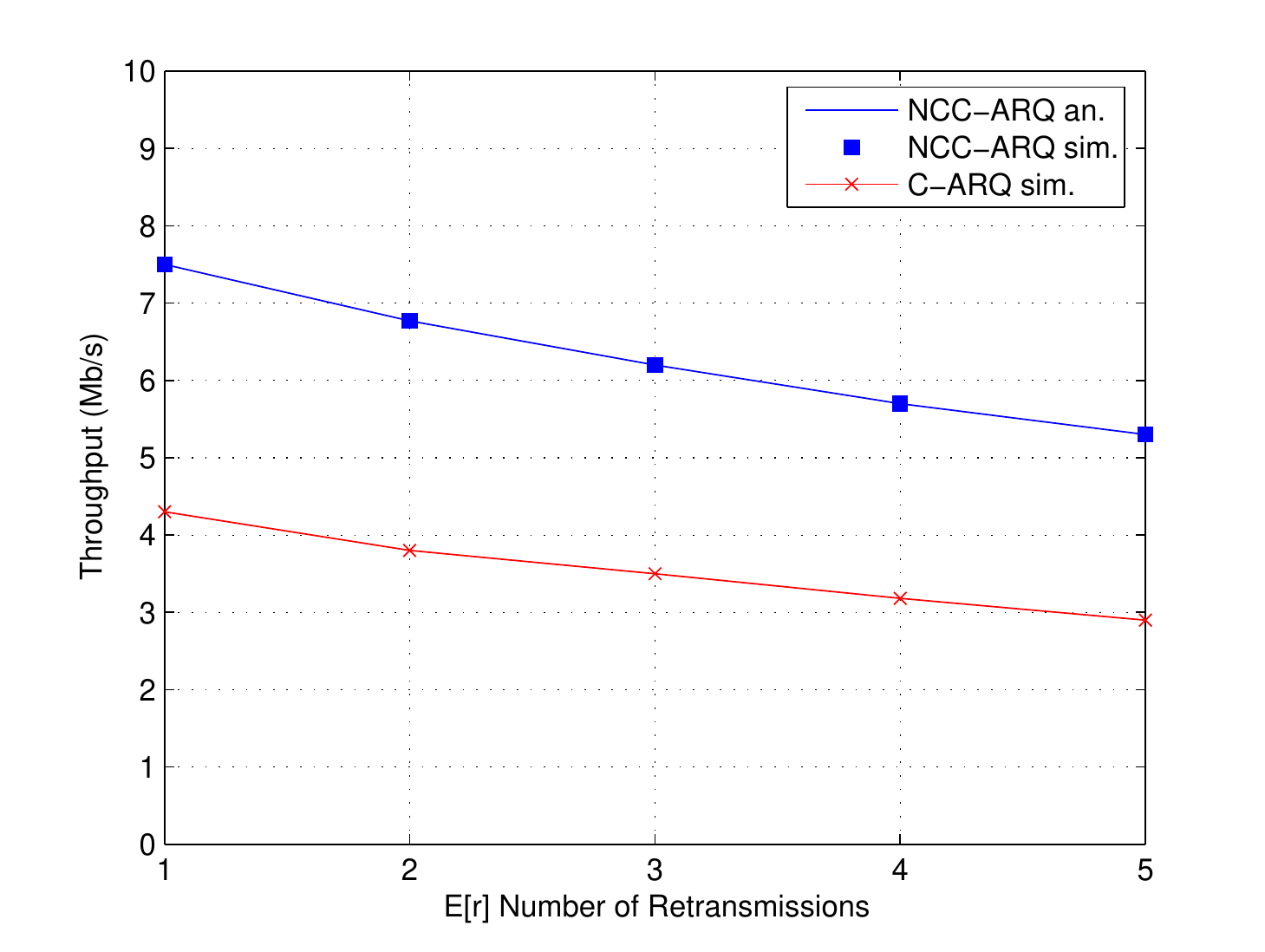}
\caption{System's Aggregated Throughput (NCC-ARQ vs C-ARQ)}\label{f4}
\end{figure}

Figure \ref{f4} shows that the numerical and the simulation results are perfectly matched, thus verifying our analysis. In addition, comparing with simple cooperative schemes which have the advantage of spatial diversity through relays without any network coding capability, we can achieve an enhancement in the network's performance up to 85\% in terms of throughput. We can see that the throughput in NCC-ARQ for one retransmission (the minimum number when the initial transmission contains errors) is 7.52 Mb/s while in simple cooperative schemes the throughput is approximately 4.3 Mb/s. Upon the increase in the number of required retransmissions (x-axe), the throughput gain is decreased, remaining though at high levels (75-85\%). This significant improvement makes sense since the number of total transmissions in NCC-ARQ scheme is lower compared to C-ARQ, as the packets are sent coded and the number of the CFC packets is decreased as well. Furthermore, in NCC-ARQ the cooperation phase is initiated only once when the traffic is bidirectional, thus saving time compared to other cooperative schemes where the cooperation takes place upon every erroneous packet reception.

\begin{figure}[htb]
\centering
\includegraphics[scale=0.6]{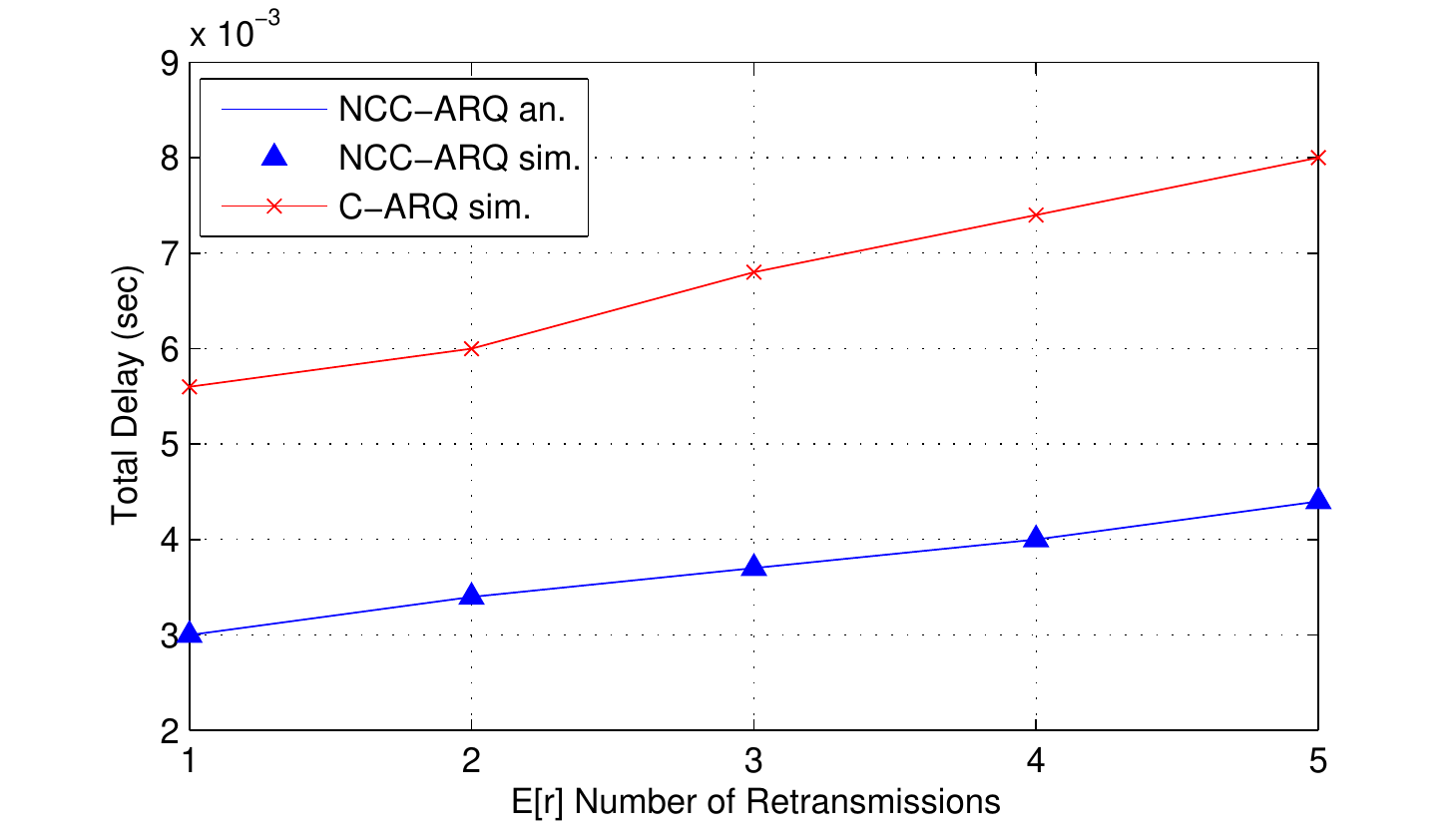}
\caption{Packet Delay (NCC-ARQ vs C-ARQ)}\label{f5}
\end{figure}

Figure \ref{f5} presents the packet delay in both Network Coding-based and simple Cooperative ARQ schemes. In this point, we must recall that two packets are delivered to their respective destinations in each transmission cycle of NCC-ARQ. Hence, in order to be accurate, we compare the delay in NCC-ARQ with the time required for two packets to be exchanged in C-ARQ.

As it can be observed, we can achieve significantly lower packet delay by using network coding techniques. Specifically, the average time that is required for two packets to be transmitted using C-ARQ is 5.6 msec in channels where one retransmission is necessary, reaching up to 8 msec when five retransmissions are required. On the other hand, the delay values in NCC-ARQ are 3 and 4.4 msec, for one and five retransmissions, respectively. This difference can be rationally explained considering the operation of our proposed NCC-ARQ scheme, since some data packets are sent to the relay (attached to the CFC message), thus avoiding the erroneous channel.

\section{Conclusion}
\label{sec:conclusion}

In this paper, a novel network coding-based cooperative ARQ (NCC-ARQ) scheme is presented. Compared to simple cooperative ARQ protocols, the proposed solution improves up to 85\% the network's aggregated bandwidth by minimizing the number of the total transmissions, while the average time to transmit data packets is significantly reduced. In order to coordinate the relay set, new MAC protocols that exploit the benefits of network coding should be designed. Our future work will be focused on such issues..

\section*{Acknowledgments}

This work has been funded by the Research Projects CO2GREEN(TEC2010-20823), GREENET(264759) and NEWCOM++(ICT-216715).

\ifCLASSOPTIONcaptionsoff
  \newpage
\fi


\begin{thebibliography}{1}

\bibitem{IEEEhowto:coop}
T. M. Cover and A. A. E. Gamal, ``Capacity Theorems for the Relay Channel", IEEE Trans. Info. Theory, vol. 25, no. 5, Sept. 1979, pp. 57284.
\bibitem{IEEEhowto:nc1}
R. Ahlswede, Ning Cai, S.-Y.R. Li, R.W. Yeung, ``Network Information Flow",  IEEE Transactions on Information Theory, vol.46, no.4, pp.1204-1216, Jul 2000
\bibitem{IEEEhowto:nc2}
S.-Y. R. Li, R. W. Yeung, Ning Cai, ``Linear network coding", IEEE Transactions on Information Theory, vol.49, no.2, pp.371-381, Feb. 2003
\bibitem{IEEEhowto:phync1}
M. Ghaderi, D. Towsley, J. Kurose, ``Reliability Gain of Network Coding in Lossy Wireless Networks", INFOCOM 2008. The 27th Conference on Computer Communications. pp.2171-2179, 13-18 April 2008
\bibitem{IEEEhowto:phync2}
R. Ahlswede, H. Aydinian, ``On Error Control Codes for Random Network Coding", Workshop on Network Coding, Theory, and Applications (NetCod) 2009. pp.68-73, 15-16 June 2009
\bibitem{IEEEhowto:phync3}
Ning Cai, ``Valuable Messages and Random Outputs of Channels in Linear Network Coding", IEEE International Symposium on Information Theory (ISIT), pp.413-417, June 28-July 3, 2009
\bibitem{IEEEhowto:phync4}
H. Bahramgiri, F. Lahouti, ``Robust Network Coding Against Path Failures", IET Communications, vol.4, no.3, pp.272-284, Feb. 12 2010
\bibitem{IEEEhowto:macnc1}
S. Katti, H. Rahul, Wenjun Hu, D. Katabi, M. Medard, J. Crowcroft, ``XORs in the Air: Practical Wireless Network Coding", IEEE/ACM Transactions on Networking , vol.16, no.3, pp.497-510, June 2008
\bibitem{IEEEhowto:macnc2}
A. Argyriou, ``Wireless Network Coding with Improved Opportunistic Listening", IEEE Transactions on Wireless Communications, vol.8, no.4, pp.2014-2023, April 2009
\bibitem{IEEEhowto:macnc3}
J. Zhang, Y. P. Chen, and I. Marsic, ``BEND: MAC-Layer Proactive Mixing Protocol for Network Coding in Multihop Wireless Networks", 9th ACM MobiHoc, Hong Kong, May 2008.

\bibitem{IEEEhowto:mr1}
K. Tan, Z. Wan, H. Zhu, J. Andrian, ``CODE: Cooperative Medium Access for Multirate Wireless Ad Hoc Network", in: Proc. of the 4th Annual IEEE Communications Society Conference on Sensor, Mesh, and Ad Hoc Communications and Networks (SECON), June 2007.

\bibitem{IEEEhowto:maccoop1}
Xin He, F. Y. Li, ``An Automatic Cooperative Retransmission MAC Protocol in Wireless Local Area Networks", European Wireless Conference (EW) 2009. pp.228-233, 17-20 May 2009
\bibitem{IEEEhowto:maccoop2}
Kejie Lu, Shengli Fu, Yi Qian, ``Increasing the Throughput of Wireless LANs Via Cooperative Retransmission", IEEE Global Telecommunications Conference (GLOBECOM), 2007. pp.5231-5235, 26-30 Nov. 2007
\bibitem{IEEEhowto:maccoop3}
J. Alonso-Zarate, E. Kartsakli, Ch. Verikoukis, and L. Alonso, ``Persistent RCSMA: A MAC Protocol for a Distributed Cooperative ARQ Scheme in Wireless Networks", EURASIP Journal on Advances in Signal Processing, vol. 2008, Article ID 817401, 13 pages, 2008. doi:10.1155/2008/817401

\bibitem{IEEEhowto:macmr1}
Pei Liu, Zhifeng Tao, S. Panwar, "A Cooperative MAC protocol for Wireless Local Area Networks," Communications, 2005. ICC 2005. 2005 IEEE International Conference on , vol.5, no., pp. 2962- 2968 Vol. 5, 16-20 May 2005
\bibitem{IEEEhowto:macmr2}
Tao Guo, R. Carrasco, ``CRBAR: Cooperative Relay-based Auto Rate MAC for Multirate Wireless Networks", IEEE Transactions on Wireless Communications, vol.8, no.12, pp.5938-5947, December 2009
\bibitem{IEEEhowto:macmr3}
Xing-Jian Zhu, Geng-Sheng Kuo, ``Cooperative MAC Scheme for Multi-Hop Multi-Channel Wireless Mesh Networks," Vehicular Technology Conference (VTC) 2008-Fall. IEEE 68th , pp.1-6, 21-24 Sept. 2008
\bibitem{IEEEhowto:macmr4}
Chang-Yeong Oh, Tae-Jin Lee, ``MAC Protocol using Cooperative Active Relays in Multi-Rate Wireless LANs", IFIP International Conference on Wireless and Optical Communications Networks (WOCN) 2009. pp.1-6, 28-30 April 2009

\bibitem{IEEEhowto:arq}
S. Lin and D.J. Costello, ``Error Control Coding: Fundamentals and Applications", Prentice-Hall, Englewood Cliffs, NJ, 1983.

\bibitem{IEEEhowto:am}
N. Morinaga, M. Nakagawa, and R. Kohno, ``New Concepts and Technologies for Achieving Highly Reliable and High-Capacity Multimedia Wireless Communications Systems", IEEE Commun. Mag., Jan. 1997, pp. 34-40.

\bibitem{IEEEhowto:phycoopnc1}
Y. Chen, S. Kishore, and J. Li, ``Wireless Diversity Through Network Coding", in Proc. WCNC, 2006, vol. 3, pp. 1681-1686.
\bibitem{IEEEhowto:phycoopnc2}
T. Wang and G. B. Giannakis, ``Complex Field Network Coding for Multiuser Cooperative Communications", IEEE Journal on Selected Areas in Communications, vol. 26, pp. 561-571, April 2008.
\bibitem{IEEEhowto:phycoopnc3}
L. Xiao, T. Fuja, J. Kliewer and D. Costello, ``A Network Coding Approach to Cooperative Diversity", IEEE Transactions on Information Theory, vol. 53, pp. 3714-3722, October 2007.

\bibitem{IEEEhowto:selection}
JiangBo Si, Zan Li and ZengJi Liu, ``Threshold Based Relay Selection Protocol for Wireless Relay Networks with Interference", IEEE International Conference on Communications (ICC), pp.1-5, 23-27 May 2010.

\bibitem{IEEEhowto:80211g}
IEEE 802.11g WG, Part 11: Wireless LAN Medium Access Control (MAC) and Physical Layer (PHY) specifications - Amendment 4: Further Higher Data Rate Extension in the 2.4 GHz Band, June 2003.


\end{thebibliography}
\end{document}